\documentclass[aps,preprint,preprintnumbers,amsmath,amssymb,floatfix]{revtex4}
\usepackage{epsfig}
\usepackage{graphicx}
\usepackage{bm}
\usepackage{amssymb}
\usepackage{dcolumn}
\usepackage{bm}
\begin{document}
\draft

\title
{Motion-selective coherent population trapping by Raman sideband cooling along two paths in a $\Lambda$ configuration}\author{Sooyoung Park, Meung Ho Seo, Ryun Ah Kim, and D. Cho\footnote{e-mail address:{\tt cho@korea.ac.kr}}}
\affiliation{Department of Physics, Korea University, Seoul 02841, Korea}

\date{\today}

\begin{abstract}
We report our experiment on sideband cooling with two Raman transitions in a $\Lambda$ configuration 
that allows selective coherent population trapping (CPT) of the motional ground state.
The cooling method is applied to $^{87}$Rb atoms in a circularly-polarized one-dimensional optical lattice.
Owing to the vector polarizability, 
the vibration frequency of a trapped atom depends on its Zeeman quantum number, 
and CPT resonance for a pair of bound states in the $\Lambda$ configuration
depends on their vibrational quantum numbers.
We call this scheme motion-selective coherent population trapping (MSCPT)
and it is a trapped-atom analogue to the velocity-selective CPT developed for free He atoms.
We observe a pronounced dip in temperature near a detuning 
for the Raman beams to satisfy the CPT resonance condition for the motional ground state.
Although the lowest temperature we obtain is ten times the recoil limit
owing to the large Lamb-Dicke parameter of 2.3 in our apparatus, 
the experiment demonstrates that MSCPT enhances the effectiveness of Raman sideband cooling
and enlarges the range of its application. 
Discussions on design parameters optimized for MSCPT on $^{87}$Rb atoms
and opportunities provided by diatomic polar molecules, 
whose Stark shift shows strong dependence on the rotational quantum number,
are included.
	
\end{abstract}

\maketitle

\section{INTRODUCTION}
Raman sideband cooling (RSC) is by far the most effective method to laser cool trapped atoms. 
The method can put atoms into the motional ground state with high probability
when the Lamb-Dicke condition $\eta_{LD}^2 =  \mathcal{E}_R /\hbar\nu\ll 1$ is satisfied.
Here, $\eta_{LD}$ is the Lamb-Dicke parameter, $\nu$ is the vibrational frequency of a trapped atom, 
and $\mathcal{E}_R$ is the recoil energy accompanying an emission of a photon. 
RSC was originally developed for an ion in a Paul trap \cite{Wineland1989}
where Coulomb interaction makes the Lamb-Dicke condition well satisfied. 
However, it is difficult to satisfy the condition in an optical dipole trap, 
and only a lattice configuration with a submicron confinement has $\eta_{LD} \ll 1$.
Early experiments applied RSC to cesium atoms in an optical lattice as a precooling stage
aiming at quantum degeneracy \cite{Weiss2000, Chu2000}.
Recently, rubidium atoms were cooled by RSC to the quantum degeneracy 
in a dynamically controlled optical lattice without resorting to evaporative cooling \cite{Vuletic2017}. 
A single atom in an optical tweezer was also cooled to the motional ground state by employing 
either a very tight focusing \cite{Regal2012} 
or an RSC sequence tailored to address high-order sidebands \cite{Ni2018}.

The order $\Delta n$ of the red sideband used in RSC should be larger than a few times $\eta_{LD}^2$ 
for the energy deficit by the red-detuned transition 
to exceed the recoil heating during an optical pumping (OP) cycle.
When $\eta_{LD} \ll 1$, tuning the Raman transition to the first-order red sideband achieves the net cooling, 
and the $n=0$ state is the only dark state, where atoms accumulate. 
Here, $n$ is the vibrational quantum number.
However, as $\eta_{LD}$ increases, $\Delta n$ should also be increased, and the $n=0$ state is no longer the only dark state.
Atoms distribute over larger $n$ raising temperature above the recoil limit $T_R = \mathcal{E}_R/k_B$.
$k_B$ is the Boltzmann constant.
In this paper, we introduce a cooling method 
that incorporates $n$-selective coherent population trapping (CPT) to RSC
so that the motional ground state remains the only dark state even outside the Lamb-Dicke regime.
We call this method motion-selective coherent population trapping (MSCPT),
and it is a trapped-atom analogue to the velocity-selective coherent population trapping 
developed for subrecoil cooling of free He atoms \cite{VSCPT 1988}. 
We apply the method to $^{87}$Rb atoms in a 1D optical lattice.
Although we cannot reach the subrecoil temperature
because we use an existing apparatus, which is not optimal for MSCPT in a few regards, 
we achieve temperature of $10T_R$ for the transverse motion, where $\eta_{LD} =2.3$,
and demonstrate that the method enhances the effectiveness of RSC and enlarges the range of its application. 

\begin{figure}[b] \centering
	\includegraphics[scale=0.35]{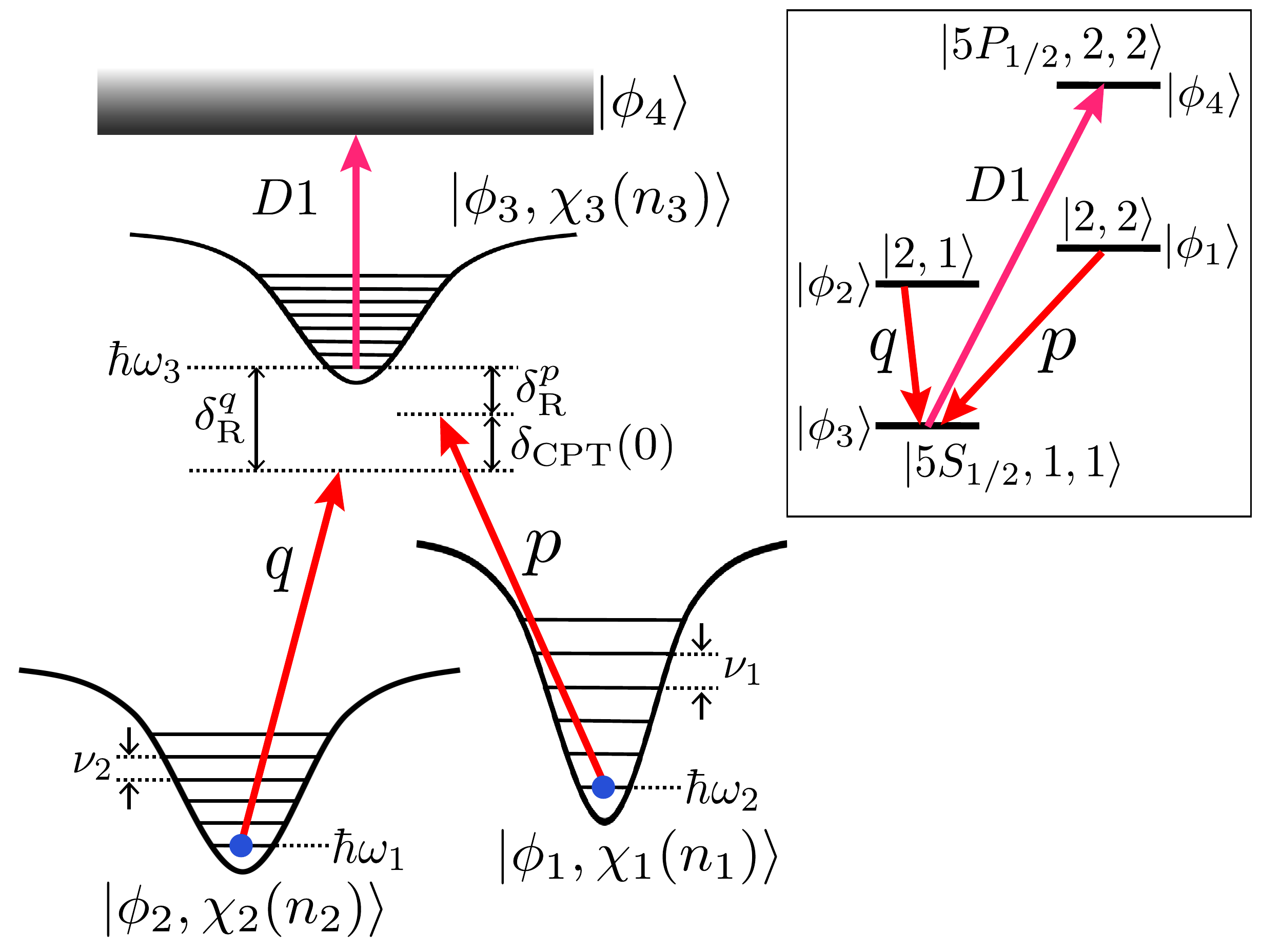}
	\caption   {\baselineskip 3.5ex 
		Inverted $\mathsf{Y}$ configuration for MSCPT experiment. 
		Specific levels and transitions in $^{87}$Rb are assigned in the inset.}
\end{figure}

\section{Theory}
Figure 1 shows the inverted $\mathsf{Y}$ configuration for our MSCPT experiment.
In the inset, specific levels and transitions in $^{87}$Rb are assigned: 
$|\phi_1 \rangle = |5S_{1/2}, F= 2, m_F = 2 \rangle$, 
$|\phi_2 \rangle =|2, 1 \rangle$,
$ |\phi_3 \rangle= |1, 1 \rangle$,
and $|\phi_4 \rangle=|5P_{1/2}, 2, 2 \rangle$.
$F$ is the total angular momentum and $m_F$ is its $z$ component. 
A usual RSC consists of a red-detuned Raman transition $|\phi_1 \rangle \rightarrow$ $ |\phi_3 \rangle$,
which we call $p$ transition,  and 
the $D$ transition $|\phi_3 \rangle \rightarrow |\phi_4 \rangle$ for an OP back to $|\phi_1 \rangle$.
To incorporate $n$-selective CPT,
we add $q$ Raman transition $|\phi_2 \rangle \rightarrow |\phi_3 \rangle$ 
to form a $\Lambda$ configuration, 
and use a circularly polarized trap beam so that $\nu_1 \neq \nu_2$
owing to the vector polarizability.
Here, $\nu_1$ and $\nu_2$ are the vibration frequencies of the motional states,
$|\chi_1\rangle$ and $|\chi_2 \rangle$, 
for the potential wells of $|\phi_1 \rangle$ and $|\phi_2 \rangle$, respectively.
When $p$ and $q$ pairs of Raman beams are tuned to 
the $\Lambda$ resonance between the $|\phi_1, \chi_1(0) \rangle$ and $|\phi_2, \chi_2(0) \rangle$ states,
the motional ground states form a CPT dark state.
A pair of $|\phi_1, \chi_1(n) \rangle$ and $|\phi_2, \chi_2(n) \rangle$ states, in general,
are detuned from the CPT resonance by $n\Delta\nu_{12}$,
where $\Delta\nu_{12}= \nu_ 1- \nu_2$, and a pair with the larger $n$ is brighter.
Atoms accumulate in the low-$n$ states, which are comparatively darker. 
$\Delta\nu_{12}$ in a circularly polarized trap is 
\begin{equation}
	\Delta\nu_{12} =\frac{\beta}{4\alpha}\,\nu_0,
	\label{eq: Delta nu 12}
\end{equation}
where $\alpha$ and $\beta$ are the scalar and vector polarizabilities, respectively, 
and $\nu_0$ is the vibration frequency in a linearly polarized trap \cite{optical Ster-Gerlach}. 
Heavy alkali-metal atoms with large $\beta$ are favored for MSCPT.
For $^{87}$Rb atom, if the trap wavelength is 860 nm and $\eta_{LD}=1$,
$\Delta\nu_{12}/ 2\pi = 65$ Hz, 
while the full width at half maximum (FWHM) of the CPT resonance 
in our radio frequency (rf) experiment was 150 Hz \cite{M1 CPT}.
In our recent publication \cite{MSCPT theory}, 
we write the master equations that describe the MSCPT scheme in Fig. 1,
and solve them numerically to show that MSCPT performs better than RSC,
and even at $\eta_{LD} \simeq 1$ it can cool atoms below $T_R$
under a favorable, but experimentally feasible, condition.

Another advantage of the inverted $\mathsf{Y}$ configuration is that it is a closed system.
The $|\phi_4 \rangle$ state decays only to one of 
the $|\phi_1 \rangle$, $|\phi_2 \rangle$, and $|\phi_3 \rangle$ states
with the probability $p_1 = 1/3,$ $p_2 =1/6,$ and $p_3 = 1/2$, respectively.
While the RSC scheme of $|\phi_1\rangle \rightarrow |\phi_3\rangle \rightarrow |\phi_4\rangle$ 
requires repumping from $|\phi_2 \rangle$
and, on average, three OP cycles to pump an atom from $|\phi_3 \rangle$ to $|\phi_1 \rangle$,
MSCPT in the inverted $\mathsf{Y}$ does not require repumping, and two OP cycles are sufficient.  
Each OP cycle, consisting of an absorption and a spontaneous emission, causes a recoil heating of $2\mathcal{E}_R$.
The $D1$ transition is used to avoid an off-resonant transition and 
subsequent decay to a state outside the $\Lambda$.
In our previous work \cite{M1 CPT}, 
using optically trapped $^7$Li atoms and rf fields for $p$ and $q$ transitions,
we demonstrated that the inverted $\mathsf{Y}$
exhibited CPT phenomena of a closed $\Lambda$ system in a wide range of experimental parameters. 

There are a few challenges to experimentally realize the idea of MSCPT.
The decoherence rate $\gamma_{12}$ between the $|\phi_1, \chi_1 \rangle$ 
and $|\phi_2, \chi_2 \rangle$ states should be minimized.
Noise in a magnetic field and phase of the Raman beams are the main sources,
and experimental techniques normally reserved for precision spectroscopy are required. 
A high-density atomic sample is not suitable for MSCPT because
the Raman and the OP beams, which stay on throughout the cooling process,
will cause photo-associative loss of atoms.
In addition, collisions between the atoms contribute to $\gamma_{12}$ \cite{collisional dephasing}.
Although the experiment we report here is carried out using a medium-density sample, 
MSCPT is best suited for a single atom in an optical lattice or an optical tweezer.
Cooling by MSCPT also proceeds slower than that by RSC because 
even a superposition of $n \neq 0$ states is partially dark, reducing the Raman transition rate.
There can also be a parasitic CPT dark state with a large $n$.

\section{Apparatus}
We use a double magneto-optical trap (MOT) system 
with an octagonal glass chamber for the second MOT.
A pair of spherical mirrors, engraved on diagonal windows of the octagon, form a Fabry-Perot cavity
for a 1D optical lattice \cite{PMSW RSC}. 
At the design wavelength $\lambda_{\rm OL}$ of 980 nm, 
FWHM of the cavity resonance is 13 MHz and the minimum spot size $w_0$ is 50 $\mu$m.
Optical mounts, coils, and a 6.8-GHz rf antenna surround the octagon,
and a double-layer magnetic shield is installed.
Fluorescence from atoms is imaged to an electron-multiplying charge-coupled device camera
with a numerical aperture of 0.28 and a threefold magnification.

\begin{figure}[b] \centering
	\includegraphics[scale=0.2]{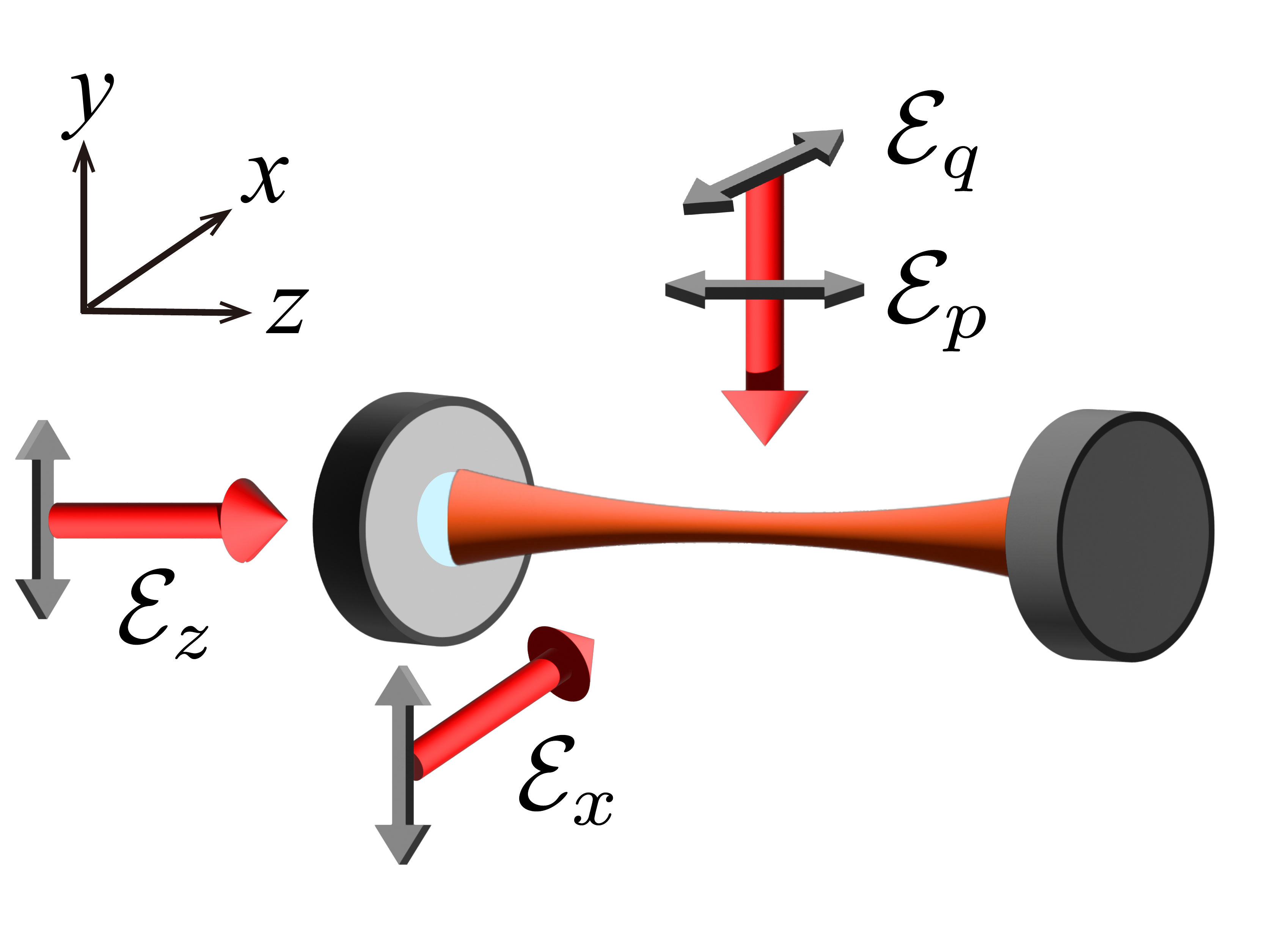}
	\caption   {\baselineskip 3.5ex 
		Raman beams applied to atoms in a 1D optical lattice for the MSCPT scheme in Fig. 1. 
		For RSC in the $xy$ plane, $(\mathcal{E}_x, \mathcal{E}_p )$ and $(\mathcal{E}_x, \mathcal{E}_q)$ pair 
		drive $p$ and $q$ transitions, respectively.  
		The $\mathcal{E}_z$ beam plays the same role as the $\mathcal{E}_x$ in the $yz$ plane.}
\end{figure}

Each of the $p$ and $q$ Raman transitions in Fig. 1 requires two laser beams, 
and their arrangement is shown in Fig. 2.
The $\mathcal{E}_x$ beam propagating along $\hat{x}$ and polarized along $\hat{y}$ pairs with
each of orthogonally polarized $\mathcal{E}_p$ and $\mathcal{E}_q$ applied along $\hat{y}$ to drive 
$p$ and $q$ transitions, respectively. 
In this way, linear momentum transferred to an atom by a $p$ transition  
is the same as that by a $q$ transition, neglecting the Zeeman shift of less than 1 MHz 
between $|\phi_1\rangle$ and $|\phi_2 \rangle$.
Otherwise, $|\phi_1, \chi_1(n) \rangle$ and $|\phi_2, \chi_2(n) \rangle$ make 
$p$ and $q$ transitions, respectively, to
groups of $|\phi_3, \chi_3(n_3) \rangle$ with different $n_3$, 
and destructive interference of the transition amplitudes for CPT is incomplete \cite{MSCPT theory}.
The Raman beams are linearly polarized so that they do not cause a differential ac Stark shift between 
the $|\phi_1\rangle$ and $|\phi_2\rangle$ states. 
The $\mathcal{E}_x$ beam is blue detuned from the $D2$ transition by 30 GHz, 
and the $\mathcal{E}_p$ and $\mathcal{E}_q$ beams are derived from it by using a fiber electro-optic modulator operating at 6.8 GHz and two acousto-optic modulators \cite{PMSW RSC}.
Oscillators driving the modulators are phase locked to an atomic clock. 
Furthermore, we add a phase-lock loop for beating between $\mathcal{E}_p$ and $\mathcal{E}_q$,
obtained immediately before the two beams enter the chamber,
because phase stability between them is critical in reducing $\gamma_{12}$.
For 3D cooling, we add an $\mathcal{E}_z$ beam which plays the same role as $\mathcal{E}_x$ in the $yz$ plane.
The $D1$ beam is added to complete the inverted $\mathsf{Y}$ configuration. 
We take much care to control its polarization to avoid an unintended transition out of $|\phi_3\rangle$ 
and subsequent decay outside $\Lambda$.
We use its intensity to control the effective decay rate $R_{\rm OP}$ of $|\phi_3 \rangle$, and hence, 
the width of the Raman transitions.

\section{Experiment and results}
Atoms are loaded into the 1D lattice and its depth is lowered to $U_0/k_B= 125$ $\mu$K. 
At this depth, inhomogeneous broadening from the vector polarizability cancels that from anharmonicity
for the red (blue) sideband of $z$ motion when the lattice beam is left (right) circularly polarized \cite{PMSW RSC}.
It narrows the sideband to facilitate RSC along the $z$ axis. 
At $U_0/k_B = 125$ $\mu$K and $w_0 = 50$ $\mu$m, along the transverse direction,
$\nu_0/ 2\pi$ = 700 Hz and $\eta_{LD} = 2.3$, being far from the Lamb-Dicke regime.
At $\lambda_{\rm OL}$ = 980 nm, $\alpha =-873$ and $\beta = -25$ in atomic units  for $^{87}$Rb,
and $\Delta \nu_{12}/2\pi = 5$ Hz from Eq.  (\ref{eq: Delta nu 12}).
Longitudinally, 
$\nu_0^{L}/2\pi = 155$ kHz, $\eta_{LD}^L = 0.15$, and $\Delta \nu_{12}^L/2\pi = 1.1$ kHz.
Atoms in the lattice show a Gaussian distribution with 100 atoms per site at the center 
and the standard deviation of 300 sites.
Their temperature is 20 $\mu$K.

As a preliminary experiment, 
we apply RSC in the transverse plane while changing the Raman detunings, defined as
$\delta_{\rm R}^p = \omega_p^x -(\omega_3 - \omega_1)$ and 
$\delta_{\rm R}^q = \omega_q^x -(\omega_3 - \omega_2)$ in Fig. 1.
Here, $\omega_p^x$ ($\omega_q^x$) is the difference between frequencies of 
the $\mathcal{E}_p$ ($\mathcal{E}_q$) and $\mathcal{E}_x$ beams in Fig. 2,
and $\hbar \omega_j$ is the energy of the $|\phi_j, \chi_j(0) \rangle$ state for $j=1,2,3$.
In this experiment, we keep $\delta_{\rm R}^p = \delta_{\rm R}^q$ 
while detuning from the CPT resonance for the $n=0$ pair of states, 
$\delta_{\rm CPT}(0) = \delta_{\rm R}^q-\delta_{\rm R}^p$, is close to zero.
In general, CPT detuning for the $n$th pair is
\begin{equation}
	\delta_{\rm CPT} (n) = (\omega_q^x - \omega_p^x) -(\omega_1-\omega_2)-n\Delta\nu_{12}.
	\label{eq:delta CPT}
\end{equation}
$\delta_{\rm R}^{p}, \delta _{\rm R}^q,$ and hence, $\delta_{\rm CPT}(0)$ are referenced to the differences,
$\omega_3 - \omega_1$ and $\omega_3 - \omega_2$, which are measured by 6.8-GHz rf spectroscopy. 
However, in a circularly polarized lattice, the rf lineshape is asymmetric and broad \cite{rf spectroscopy},
and extracting the difference by fitting a theoretical curve to a lineshape is susceptible to errors.
We estimate the uncertainty to be 1 kHz.
For the experiment, atoms are optically pumped to $|\phi_1 \rangle$, and  
$\mathcal{E}_p$, $\mathcal{E}_q$, $\mathcal{E}_y$ and the OP beams are applied.
The quantization field is 1 G,
the Rabi frequencies are $\Omega_p/2\pi = 4.3$ kHz and $\Omega_q/2\pi = 7.6$ kHz for $p$ and $q$ transitions, respectively, and $R_{\rm OP}/2\pi = 2.8$ kHz.
After the beams are applied for 3 s, only 20\% of the atoms are left mainly owing to the photo-associative loss.
The transverse temperature $T$ is measured by the time-of-flight method.
Red squares in Fig. 3 denote $T$ versus the common detuning $\delta_{\rm R}$.
$T$ increases sharply near $\delta_{\rm R}^\prime=\delta_{\rm R}/2\pi = -15$ kHz, 
which is the minimum red detuning to overcome the recoil heating of $4\mathcal{E}_R$. 
The lowest $T$ of 4.7 $\mu$K is obtained at $\delta_{\rm R}^\prime = -60$ kHz,
limited by the cross-dimensional heating by the $z$ motion.
We add the $\mathcal{E}_z$ beam, tuned to the red sideband of the $z$ motion, and 
obtain the lowest $T$ of 3.0 $\mu$K at $\delta_{\rm R}^\prime = -30$ kHz, as represented by blue circles in Fig. 3.
Removing the heat load from the $z$ motion allows us to use smaller $|\delta_{\rm R}|$,
which, in turn, narrows the distribution of the atoms over $n$.
The cooling time required to reach the steady state is also reduced to 2 s.
Although addition of the $\mathcal{E}_z$ beam adds another $\Lambda$ configuration,
it does not interfere with MSCPT in the $xy$ plane 
because, once an atom falls into the $n_z =0$ state, transitions driven by $\mathcal{E}_z$ are far-off resonant. 

\begin{figure}[t] \centering
	\includegraphics[scale=0.43]{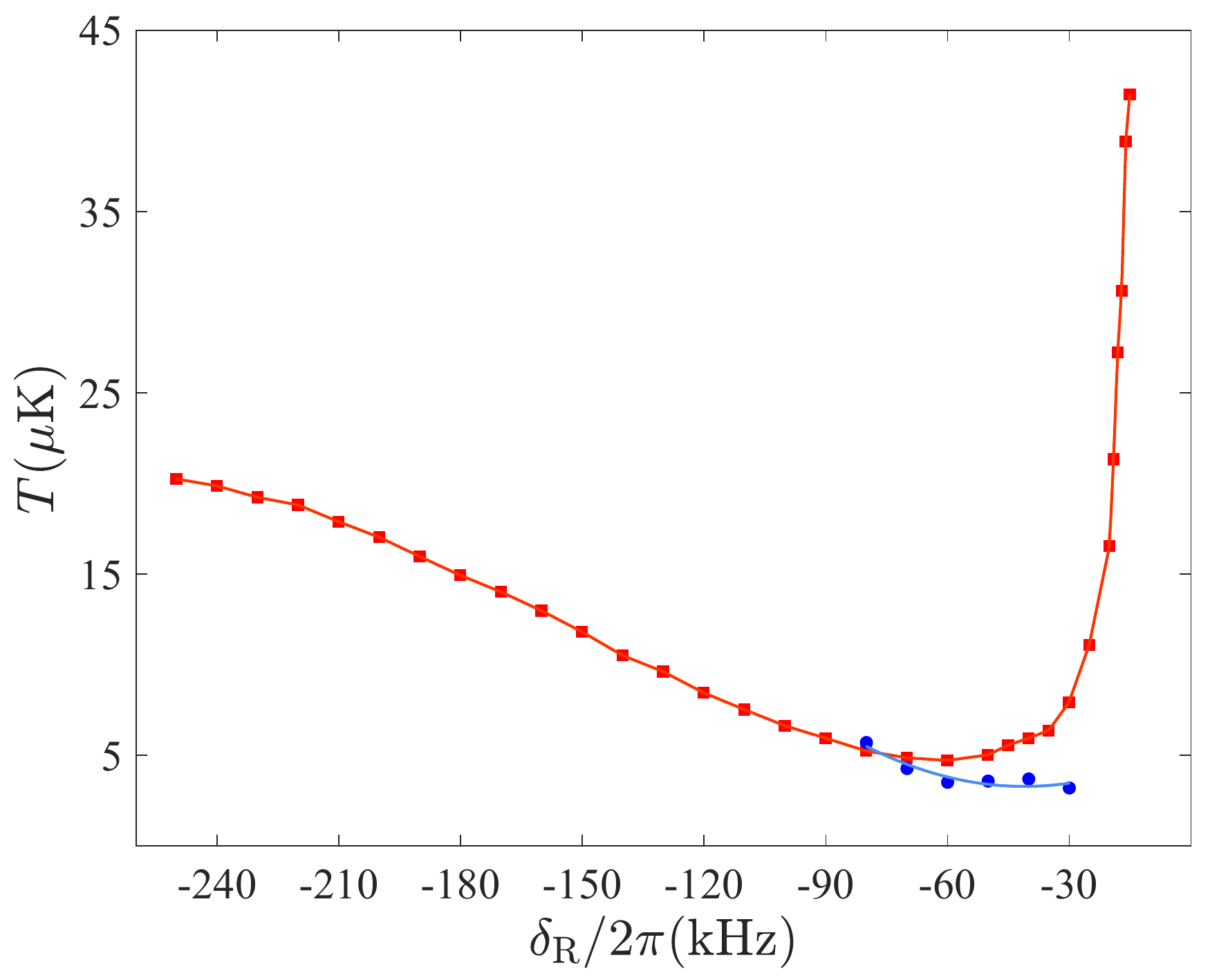}
	\caption   {\baselineskip 3.5ex 
		Temperature of transverse motion versus the common detuning $\delta_{\rm R}$ of $p$ and $q$ Raman transitions. 
		Detuning from the CPT resonance for a pair of motional ground states, $\delta_{\rm CPT}(0)$,  is kept close to zero. 
		Red squares are from 2D RSC in the $xy$ plane and 
		blue circles are from 3D RSC with the addition of the $\mathcal{E}_z$ beam.}
\end{figure}

\begin{figure}[t] \centering
	\includegraphics[scale=0.43]{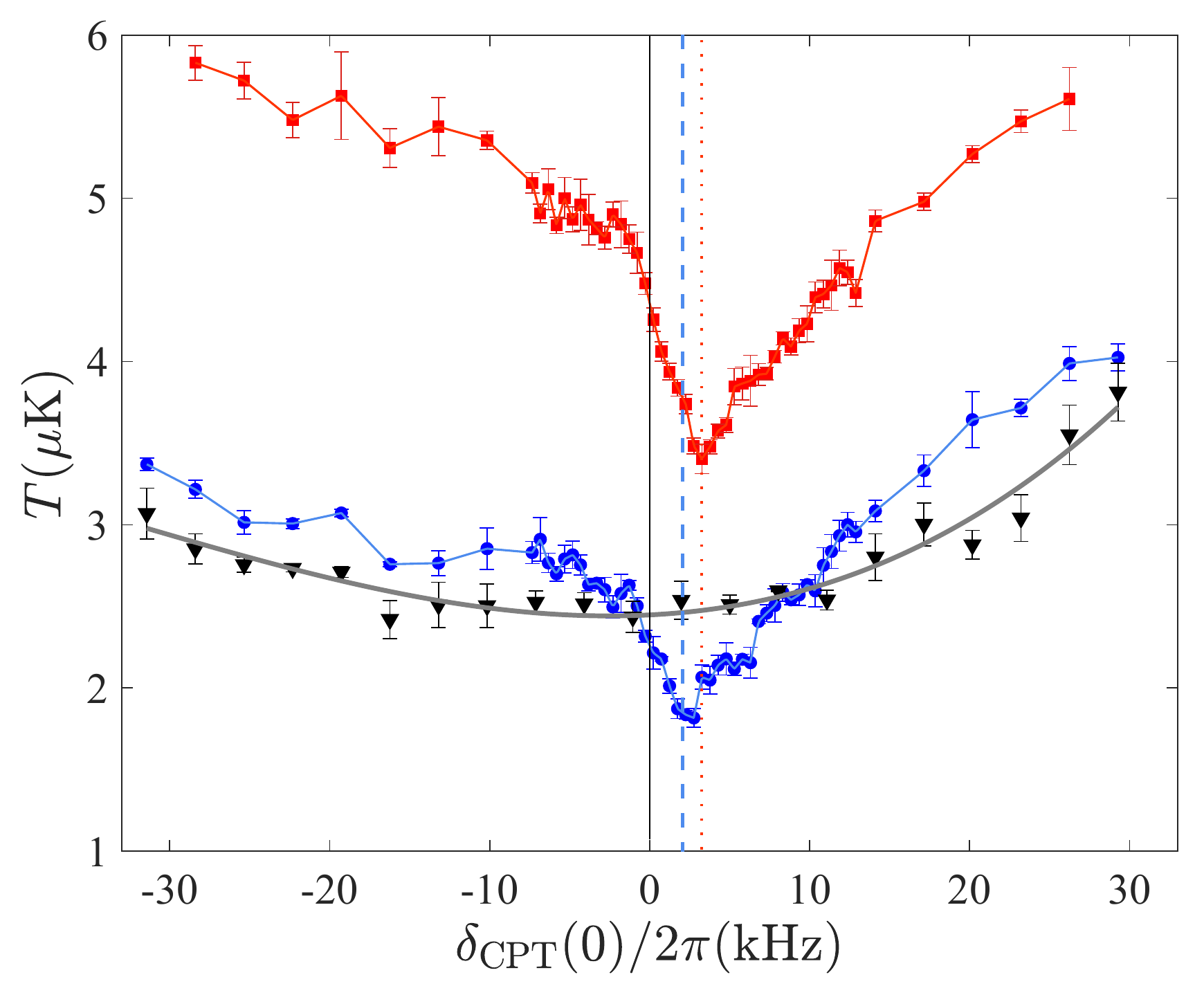}
	\caption   {\baselineskip 3.5ex 
		Temperature of transverse motion versus detuning from the CPT resonance for the motional ground state,
		$\delta_{\rm CPT}(0) = \delta_{\rm R}^q - \delta_{\rm R}^p$.
		$\delta_{\rm R}^q$ is scanned while $\delta_{\rm R}^p/2\pi$ is kept at -60 kHz in 2D MSCPT (red square), 
		and at -30 kHz in 3D MSCPT (blue circle). 
		Red dotted and blue dashed vertical lines denote $\delta_{\rm CPT}(0)$ where the lowest temperature in 2D and 3D experiment occurs, respectively.
		Black triangles represent results when the $\mathcal{E}_p$ and $\mathcal{E}_q$ beams are alternately turned on at 1 kHz.}
\end{figure}

As we scan $\delta_{\rm CPT}(0) = \delta _{\rm R}^q -\delta_{\rm R}^p$ by changing $\delta_{\rm R}^q$,
a pronounced dip in $T$ appears, demonstrating the extra cooling effect of MSCPT.
Figure 4 shows the results from an experiment in 2D at 
$\delta^{\prime \, p}_{\rm R} =\delta_{\rm R}^p /2\pi =  -60$ kHz (red square)
and that in 3D at $\delta^{\prime \, p}_{\rm R} = -30$ kHz (blue circle).
For a direct comparison with RSC in 3D, 
we alternately turn on the $\mathcal{E}_p$ and $\mathcal {E}_q$ beams at 1 kHz for 4 s 
while $\mathcal{E}_x$ and $\mathcal {E}_z$ stay on. 
The results (black triangle) show gradual change in $T$ without a dip. 
In the 2D MSCPT experiment, 
the lowest $T$ of 3.4 $\mu$K occurs at $\delta^\prime_{\rm CPT}(0) =\delta_{\rm CPT}(0)/2\pi =  3.3$ kHz, 
and in 3D, $ T= 1.8$ $\mu$K at $\delta^\prime_{\rm CPT}(0) = 2.2$ kHz.
Ideally, the lowest $T$ should be at $\delta_{\rm CPT}(0) = 0$, and
we interpret this shift as a result of the width of the Raman transition being much smaller than $|\delta_R|$.
The half-width at half maximum, contributed by the power and radiative broadening, for the $p$ and $q$ transition
is 3.3 kHz and 5.5 kHz, respectively.
The low-$n$ states are already dark owing to the much larger Raman detuning, 
and the CPT does not have an impact on the atoms in those states.
From Eq. (\ref{eq:delta CPT}), at positive $\delta_{\rm CPT}(0)$,
$\omega_q^x - \omega_p^x$ is CPT resonant for the $n >0$ states,
and we conjecture that the observed position of the dip is
where the CPT-induced darkness can best complement that induced by $|\delta_{\rm R}|$
to produce the lowest $T$.
$\delta_{\rm CPT}(0)$ at the dip in 2D with $\delta^{\prime \, p}_{\rm R} = -60$ kHz is one and a half times 
that  in 3D with $\delta^{\prime \, p}_{\rm R}= -30$ kHz,
and it is consistent with the conjecture. 
The 1D simulations also show that MSCPT loses advantage over RSC 
when the Raman width is much smaller than $|\delta_{\rm R}|$ \cite{MSCPT theory}.
In our experiment, the available laser power and the loss of atoms from photo-association 
limit $\Omega_{p,q}$ and $R_{\rm OP}$, and hence, the range of the Raman width. 
In addition to being the result of an extra cooling effect, the dips are, to the best of our knowledge, 
the first observation of CPT phenomena driven by a pair of stimulated Raman transitions.
To confirm this, we repeat $\delta_{\rm R}^q$ scans at a few $\delta_{\rm R}^p$.
The results are shown in Fig. 5: three upper traces from the right to the left are 
at $\delta^{\prime \, p}_{\rm R} = -30, -40,$ and -60 kHz in 2D,
and two lower traces are at $\delta^{\prime \, p}_{\rm R} = -30$ and -60 kHz in 3D. 
The dips appear when $\delta_{\rm R}^q \simeq \delta_{\rm R}^p$, 
independent of $\delta_{\rm R}^q$ and $\delta_{\rm R}^p$ themselves. 

\begin{figure}[t] \centering
	\includegraphics[scale=0.42]{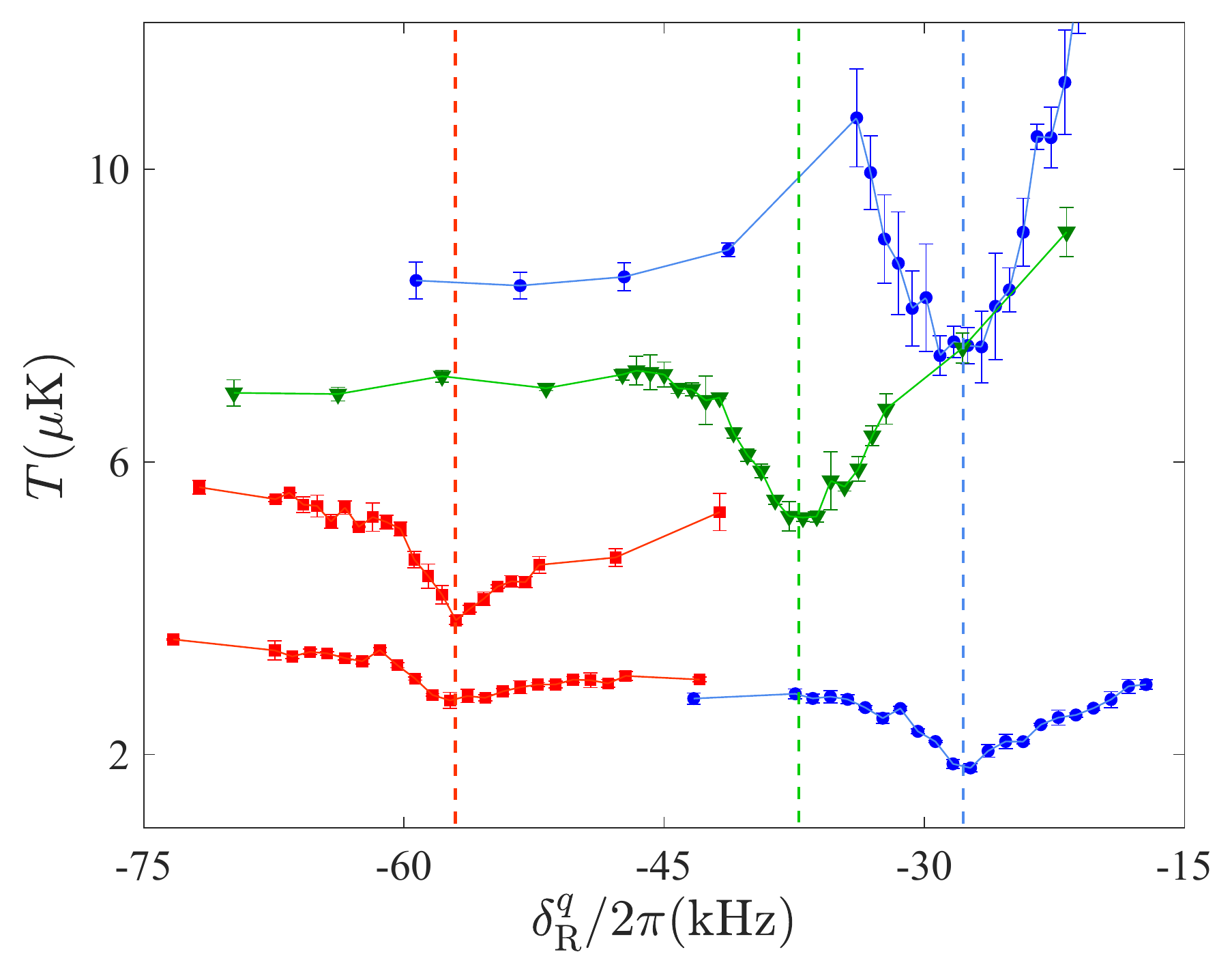}
	\caption   {\baselineskip 3.5ex 
		Temperature of transverse motion versus $\delta_{\rm R}^q$ at a few $\delta_{\rm R}^p$.
		Three upper traces are from 2D MSCPT at $\delta_{\rm R}^p/2\pi =$ 
		-30 (blue circle), -40 (green triangle), and -60 kHz (red square).
		Two lower traces are form 3D MSCPT at $\delta_{\rm R}^p/2\pi =$ -30 and -60 kHz.
	    Vertical dashed lines are placed where temperature is the lowest for each $\delta_{\rm R}^q$ scan.}
\end{figure}

In MSCPT, it is critical to minimize the decoherence rate $\gamma_{12}$ between $|\phi_1, \chi_1 \rangle$ and $|\phi_2, \chi_2 \rangle$.
We consider two sources for $\gamma_{12}$: 
noise $\delta B(t)$ in a magnetic field
and phase fluctuation $\delta \phi(t)$ between the $\mathcal{E}_p$ and $\mathcal{E}_q$ beams.
For a given $\delta B(t)$, $\gamma_{12}$ is proportional to $\Delta \nu_{12}^2$, 
and we use the hyperfine transition between $ |5S_{1/2}, F = 1, m_F=1 \rangle$ and $|F= 2, m_F =0\rangle$
as a substitute for the $|\phi_1 \rangle \rightarrow |\phi_2 \rangle$ transition to measure $\gamma_{12}$.
The ladder structure of the Zeeman sublevels in the $|5S_{1/2}, F=2 \rangle$ state
makes spectroscopy on the isolated $|\phi_1 \rangle$ and $|\phi_2 \rangle$ states difficult. 
In order to be free from inhomogeneous broadening, 
we use spin-echo spectroscopy on atoms in a linearly polarized lattice at the well depth of only 11 $\mu$K.
With a two-layer magnetic shield and a low-noise current supply, 
$\gamma_{12}$ from $\delta B(t)$ at $B_0 =1$ G is $2\pi \times 1.8$ Hz. 
$\gamma_{12}$ from $\delta \phi(t)$, estimated from an rf spectrum of
the beating between $\mathcal{E}_p$ and $\mathcal{E}_q$, is $2\pi \times 1.8$ Hz
and it is reduced to $2\pi \times 1.4$ Hz when the beating signal is phase locked to an atomic clock.
Figure 6(a) shows the lowest $T$ in 3D MSCPT versus $\log_{10} (\gamma_{12}/\gamma_{12}^0)$, 
where $\gamma_{12}^0 = 2\pi \times 1$ Hz. 
$\gamma_{12}$ is increased by injecting white noise to the rf which modulates the $\mathcal{E}_q$ beam. 
We observe gradual disappearance of the dip and an increase in $T$.
Finally, we change the degree of circularity of the lattice beam,
 $\zeta = i \hat{z} \cdot (\hat{\epsilon} \times \hat{\epsilon}^*)$, where $\hat{\epsilon}$ is the Jones vector.
For a given $\zeta$, 
$\Delta \nu_{12}$ in Eq. (\ref{eq: Delta nu 12}) is reduced to $\zeta(\beta/4\alpha)\nu_0$. 
Figure 6(b) shows the increase in the lowest $T$ in 3D MSCPT as the lattice polarization becomes linear. 
When the polarization is linear, a peak as well as a dip in $T$ appears as we scan $\delta_{\rm CPT} (0)$, 
indicating unintended trapping of population at high-$n$ states.

\begin{figure}[t] \centering
	\includegraphics[scale=0.4]{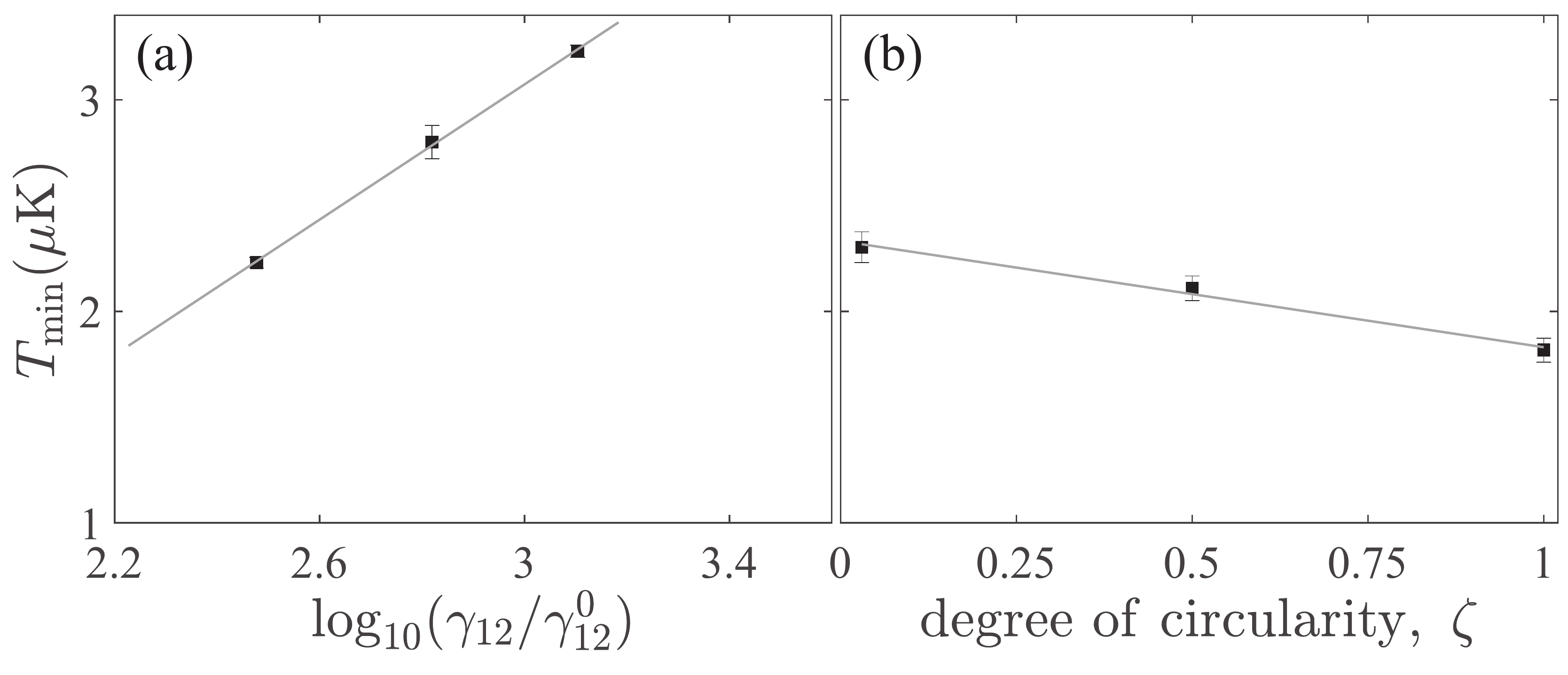}
	\caption   {\baselineskip 3.5ex 
		(a) Minimum temperature at a given $\gamma_{12}$ in 3D MSCPT versus 
		$\log_{10} (\gamma_{12}/\gamma_{12}^0)$, where $\gamma_{12}^0 = 2\pi\times 1$ Hz.
		(b) Minimum temperature at a given degree of circularity, $\zeta$, of the optical lattice beam in 3D MSCPT versus 
		$\zeta$.}
\end{figure}

\section{Summary and Discussion}
In summary, we apply Raman sideband cooling in a $\Lambda$ configuration 
to atoms in a circularly-polarized 1D optical lattice.
State-dependent variation in vibration frequency allows 
us to tune the Raman beams for coherent population trapping of the motional ground states.
Although we observe CPT dark state of the low-$n$ states only and not the $n=0$ state 
and the lowest temperature obtained is 10 times the recoil limit, 
our results clearly demonstrate $n$-selective CPT phenomena and a cooling efficiency better than that of RSC.
To realize the full benefit of the MSCPT scheme, we are building an apparatus optimized for the scheme.
The apparatus used in the experiment reported here has $\eta_{LD} = 2.3$ and $\lambda_{\rm OL} =980$ nm, 
resulting in too closely spaced vibrational levels and too small $\Delta\nu_{12}$, respectively. 
The design parameters of the present and the new apparatus are compared in Table I.
Fivefold reduction of $w_0$ at the same well depth $U_0$ reduces $\eta_{LD}$ to 1,
increasing $\nu_0/2\pi$ from 700 Hz to 3.5 kHz.
This reduces the density of states in the $xy$ plane by a factor of 25, 
decreasing the number of diffusive steps required to reach the ground state. 
At $\lambda_{\rm OL} = 860$ nm, $\beta/\alpha$ increases by a factor of 2.5.
With the new design parameters, $\Delta\nu_{12}/2\pi$ increases from 5 Hz to 65 Hz,
and we expect significant enhancement in the $n$ selectivity.
The photon scattering rate $R_\gamma$ also increases by 2.5, but it is still less than $2\pi \times 0.1$ Hz.
Finally, we note that a diatomic polar molecule in an optical trap \cite{polar molecule} provides 
an excellent opportunity to apply MSCPT
because its Stark shift depends strongly on the rotational quantum number.
Considering a MgF molecule in a 532-nm optical trap as an example, 
the fractional difference between vibration frequencies of a pair of states in the same ro-vibrational level 
can be as large as 12\%. 
This may be compared with $\Delta \nu_{12}/\nu_0$ of less than 2\% for $^{87}$Rb in the new design. 

\begin{table}[t]
	\begin{tabular} {c||c|c} \hline \hline
			& \hspace{3 mm} present apparatus  \hspace{3 mm} & \hspace{3 mm} new apparatus  \hspace{3 mm}\\ \hline
			$U_0/k_B$ ($\mu$K) & 125		& 125                         \\ \hline
			$w_0$ ($\mu$m)     & 50                          & 10 \\ \hline
			$\eta_{LD}$ & 2.3 & 1.0 \\ \hline
			$\nu_0/2\pi$ (kHz) & 0.7 & 3.5 \\ \hline
			$\lambda_{\rm OL}$ (nm) & 980 & 860 \\ \hline
			$\alpha$ (atomic unit) \hspace{3 mm} & -873 & -1893 \\ \hline
			$\beta$ (atomic unit) \hspace{3 mm} & -25 & -139 \\ \hline
			$\Delta\nu_{12}/2\pi$ (Hz)  &  5 &  65 \\  \hline
			$R_\gamma /2\pi$ (Hz)  & 0.03 &  0.07 \\ \hline \hline
		\end{tabular} 
		\caption{Design parameters of the present and the new machine for MSCPT experiment using $^{87}$Rb atoms.
			\label{Table1}}
\end{table}

\section*{ACKNOWLEDGMENTS}
This work was supported by the National Research Foundation of Korea (Grant No. 2017R1A2B3002543).


\newpage
\begin{thebibliography}{99}
\bibitem{Wineland1989}
F. Diedrich, J. C. Bergquist, W. M. Itano, and D. J. Wineland,
Laser Cooling to the Zero-Point Energy of Motion,
Phys. Rev. Lett. {\bf 62}, 403 (1989).	

\bibitem{Weiss2000}
D.-J. Han, S. Wolf, S. Oliver, C. McCormick, M. T. DePue, and D. S. Weiss,
3D Raman Sideband Cooling of Cesium Atoms at High Density,
Phys. Rev. Lett. {\bf 85}, 724 (2000).

\bibitem{Chu2000}
A. J. Kerman, V. Vuleti\'c, C. Chin, and S. Chu,
Beyond Optical Molasses: 3D Raman Sideband Cooling of Atomic Cesium to High Phase-Space Density,
Phys. Rev. Lett. {\bf 84}, 439 (2000).

\bibitem{Vuletic2017}
J. Hu, A. Urvoy, Z. Vendeiro, V. Crépel, W. Chen, V. Vuleti\'c,
Creation of a Bose-condensed gas of $^8$$^7$Rb by laser cooling,
Science {\bf 358}, 1078 (2017).

\bibitem{Regal2012}
A. M. Kaufman, B. J. Lester, and C. A. Regal,
Cooling a Single Atom in an Optical Tweezer to Its Quantum Ground State,
Phys. Rev. X {\bf 2}, 041014 (2012).

\bibitem{Ni2018}
Y. Yu, N. R. Hutzler, J. T. Zhang, L. R. Liu, J. D. Hood, T. Rosenband, and K.-K. Ni,
Motional-ground-state cooling outside the Lamb-Dicke regime, Phys. Rev. A 97, 063423
(2018).

\bibitem{VSCPT 1988}	
A. Aspect, E. Arimondo, R. Kaiser, N. Vansteenkiste, and C. Cohen-Tannoudji, 
Laser Cooling below the One-Photon Recoil Energy by Velocity-Selective Coherent Population Trapping,
Phys. Rev. Lett. {\bf 61}, 826 (1988).

\bibitem{optical Ster-Gerlach}
C. Y. Park, J. Y. Kim, J. M. Song, and D. Cho,
Optical Stern-Gerlach effect from the Zeeman-like ac Stark shift,
Phys. Rev. A {\bf 65}, 033410 (2002).

\bibitem{M1 CPT}	
H. Kim, H. S. Han, T. H. Yoon, and D. Cho,
Coherent Population Trapping in a $\Lambda$ Configuration Coupled by Magnetic Dipole Interactions,
Phys. Rev. A {\bf 89}, 032507 (2014).

\bibitem{MSCPT theory}
H. G. Lee, S. Park, M. H. Seo, and D. Cho,
Motion-selective coherent population trapping for subrecoil cooling of optically trapped atoms outside the Lamb-Dicke regime,	arXiv:2205.00685 [physics.atom-ph].


\bibitem{collisional dephasing}
Y. Sagi, I. Almog, and N. Davidson,
Universal Scaling of Collisional Spectral Narrowing in an Ensemble of Cold Atoms,
Phys. Rev. Lett. {\bf 105}, 093001 (2010).

\bibitem{PMSW RSC}
M. H. Seo, S. Park, and D. Cho,
Relaxation of atomic temperature anisotropy in a one-dimensional optical lattice enhanced 
by dynamic control of the aspect ratio,
Phys. Rev. A {\bf 101}, 043611 (2020).

\bibitem{rf spectroscopy}
S. Park, M. H. Seo, and D. Cho,
Ground-state hyperfine spectroscopy of $^{87}$Rb atoms in a 1D optical lattice,
J. Phys. B: At. Mol. Opt. Phys. {\bf{52}}, 235002 (2019).

\bibitem{polar molecule}
L. Anderegg, B. L. Augenbraun, Y. Bao, S. Burchesky, L. W. Cheuk, W. Ketterle, and J. M. Doyle,
Laser Cooling of Optically Trapped Molecules,
Nat. Phys. {\bf 14} 890 (2018).

\end{thebibliography}
\end{document}